\documentclass[manuscript, preprint]{acmart}
\AtBeginDocument{%
  }
\usepackage{subcaption}
\usepackage{booktabs}
\usepackage{multirow}
\usepackage{mathtools}
\usepackage[super]{nth}
\setcopyright{acmlicensed}
\copyrightyear{2018}
\acmYear{2018}
\acmDOI{XXXXXXX.XXXXXXX}
\acmConference[Conference acronym 'XX]{Make sure to enter the correct
  conference title from your rights confirmation email}{June 03--05,
  2018}{Woodstock, NY}
\acmISBN{978-1-4503-XXXX-X/2018/06}

\begin{document}

\title{Understanding Role Switching in Human–AI Collaboration through Multimodal Behavioral Signals}

\author{Avinash Ajit Nargund}
\authornote{Both authors contributed equally to this research.}
\email{anargund@ucsb.edu}
\author{Arthur Caetano}
\email{caetano@ucsb.edu}
\authornotemark[1]
\affiliation{%
  \institution{University of California, Santa Barbara}
  \city{Santa Barbara}
  \state{CA}
  \country{USA}
}

\author{Kevin S Yang}
\affiliation{%
  \institution{University of California, Santa Barbara}
  \city{Santa Barbara}
  \state{CA}
  \country{USA}
}

\author{Rose Yiwei Liu}
\affiliation{%
  \institution{Washington University in St.Louis}
  \city{St.Louis}
  \state{MO}
  \country{USA}
}

\author{Pranav Raghavendra Gunhal}
\affiliation{%
  \institution{University of California, Santa Barbara}
  \city{Santa Barbara}
  \state{CA}
  \country{USA}
}

\author{Philip Tezaur}
\affiliation{%
  \institution{University of California, Santa Barbara}
  \city{Santa Barbara}
  \state{CA}
  \country{USA}
}

\author{Kriteen Shrestha}
\affiliation{%
  \institution{University of California, Santa Barbara}
  \city{Santa Barbara}
  \state{CA}
  \country{USA}
}

\author{Qisen Pan}
\affiliation{%
  \institution{University of California, Santa Barbara}
  \city{Santa Barbara}
  \state{CA}
  \country{USA}
}

\author{Tobias Höllerer}
\email{holl@ucsb.edu}
\affiliation{%
  \institution{University of California, Santa Barbara}
  \city{Santa Barbara}
  \state{CA}
  \country{USA}
}

\author{Misha Sra}
\email{sra@ucsb.edu}
\affiliation{%
  \institution{University of California, Santa Barbara}
  \city{Santa Barbara}
  \state{CA}
  \country{USA}
}

\renewcommand{\shortauthors}{Trovato et al.}

\begin{abstract}
Human–AI collaboration often requires dividing complex tasks into complementary subtasks. As a task unfolds, users may want to shift which subtasks they perform and which their AI partner performs, in response to evolving task demands and perceptions of the AI's capabilities
In this work, we investigate whether behavioral signals can reveal such changes during a sequential decision-making task.
We conducted a study using hand-and-brain chess, where, on each turn, participants chose either to select the piece type (brain) while their AI partner chose the move (hand), or to choose the move after the AI selected the piece type. Across 21 chess players, this yielded more than 1,100 decisions to retain their role from the previous turn or switch to the other role.
Players generally retained their current roles across turns. When participants did switch, they exhibited more exploratory gaze patterns and role switches were associated with lower subsequent move quality. Using these behavioral and task-specific signals, we trained a classifier to distinguish switch from stay decisions, achieving a PR-AUC of 0.56 (compared to a random baseline of 0.39). Feature-set ablations showed that gaze and task-specific features contributed most to model performance.
Role switching was not a simple choice between controlling and delegating the task as both roles required participants to perform one subtask and delegate the other. However, interviews showed that many participants perceived selecting the piece type as giving them greater control. Perceived AI ability, relative subtask difficulty, and desired influence over the direction of play shaped these role preferences. These findings suggest that behavioral cues could help intelligent systems recognize when users want to reallocate complementary responsibilities during collaboration.

\end{abstract}


\begin{teaserfigure}
  \includegraphics[width=\textwidth]{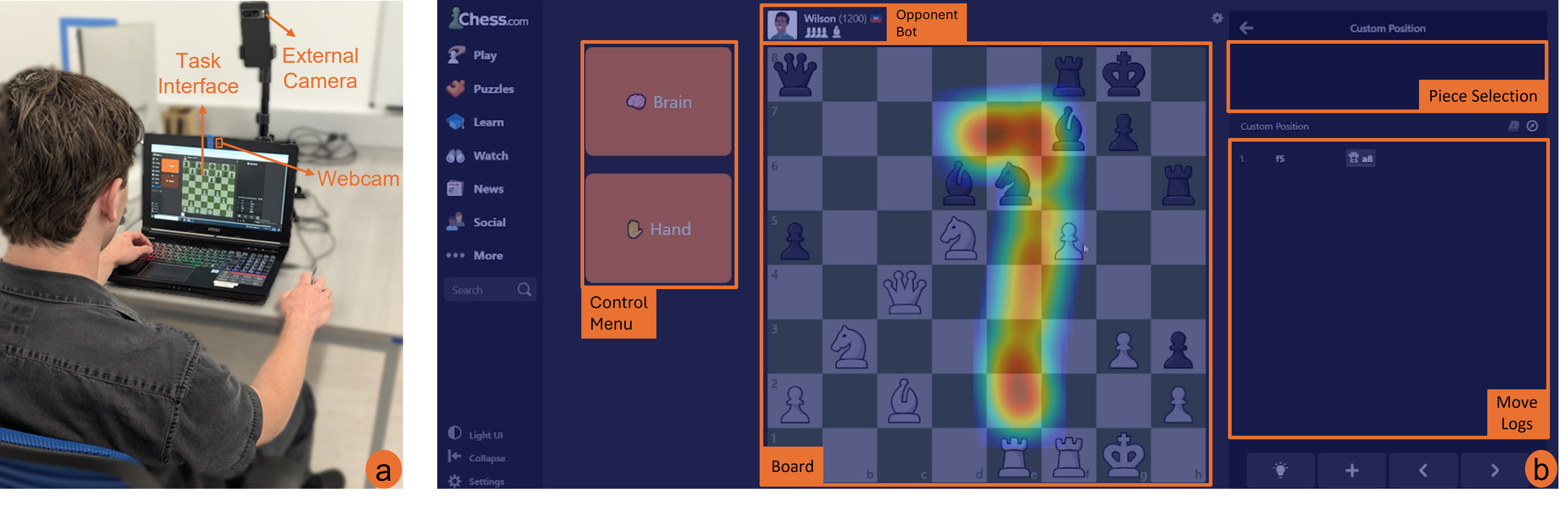}
  \caption{\textbf{Investigating Role Switching in Human–AI Collaboration.} We examine whether role switches during a sequential decision-making task can be detected using behavioral and task-based signals. Participants played modified hand-and-brain chess with an AI teammate, choosing on each turn whether to take the object-level role (selecting a piece type) or the action-level role (executing a specific move). The left panel (a) shows a participant interacting with our custom Chess.com browser-extension interface. Behavioral data, including gaze (recorded via the laptop webcam) and role selections (recorded by the browser extension), were collected. Emotional state was assessed post-hoc using facial videos recorded with an external camera. The right panel (b) shows a detailed view of the interface. An overlaid gaze map illustrates a shift in attention from the white rook on square \texttt{e1} toward the black knight on \texttt{e6}. Note that the board position shown in panel (a) is distinct from the position in panel (b).}
  \label{fig:teaser}
  \Description{The left panel of this figure shows a participant playing hand-and-brain chess with an AI teammate on a laptop using a custom interface, with gaze tracking and video recording equipment visible. The right panel is a detailed view of the interface with the role selection menu, board, move log area and piece selection area highlighted and gaze heatmap overlaid to illustrate attention shifts during gameplay.}
\end{teaserfigure}

\received{20 February 2007}
\received[revised]{12 March 2009}
\received[accepted]{5 June 2009}

\maketitle

\section{Introduction}

Artificial intelligence (AI) systems are increasingly integrated into human decision-making in domains such as medical imaging, autonomous driving, and finance. Human-AI collaboration in these domains often involves dividing complex tasks into subtasks handled by either the human or AI, with sequential decision-making where each choice shapes subsequent actions and outcomes. As tasks progress, user preferences for which subtasks to handle may shift based on evolving trust, decision complexity, and desired control. 

Despite the growing use of AI tools to support human decision-making, their integration into effective collaborative workflows remains challenging. A recent meta-analysis by ~\citet{vaccaro2024combinations} found that human-AI teams often perform worse than the better of the two working alone, particularly in decision-making tasks. One contributing factor is that users often lack the ability to fluidly shift their role as task demands change~\cite{chiou2021mixed, amershi2019guidelines}. This limitation arises because division of work between human and AI is determined through static configurations or dynamic system-driven adaptations. Static schemes define how subtasks are allocated in advance~\cite{de2008mixed}. They usually lack the flexibility to accommodate changing task demands or user behavior and can encode human biases at design time~\cite{pinski2023, fugener_cognitive_2022, glickman2025human}. System-driven approaches instead adapt task allocation using signals such as task priority, confidence, inferred user goals or competence~\cite{amershi2019guidelines, rabby2022learning, salikutluk_evaluation_2024}. Although more flexible, these adaptations might not align with how users prefer to divide responsibilities with intelligent systems, causing friction or loss of trust~\cite{tambe2000, hauptman_adapt_2023}. Mixed-initiative approaches provide a means for humans and AI systems to participate more flexibly in determining how work is divided~\cite{bradshaw2003dimensions}, but few are grounded in empirical evidence of evolving role preferences in long sequential tasks.

Prior work on task delegability has primarily examined whether a task or decision should be performed by the human or delegated to an AI~\cite{lubars2019, amershi2019guidelines}. In contrast, we examine repeated role changes within the same sequential task, where both the human and AI remain active collaborators performing complementary subtasks. This allows us to study how users revise their preferred role as the task progresses and whether those shifts can be inferred from behavioral and task-specific signals. Adopting ~\citet{lubars2019}'s distinction between object-level and action-level decisions, we study how users revise their preferred role as the task progresses and whether these shift can be inferred from behavioral and task-specific signals. This turn-based structure lets us examine how gaze, affect, and task state relate to shifts in role preference, extending research on trust and cognitive bias in human-AI collaboration~\cite{gurney2023role}.

To study these role-switching dynamics, we adapted the hand-and-brain chess variant ~\footnote{\url{https://www.chess.com/terms/hand-and-brain-chess}} into a human-AI collaboration task (See Figure~\ref{fig:teaser}). We chose chess because it provides a structured sequential task with clearly separable, complementary subtasks, repeated opportunities for role changes, and objective measures of task state and performance. On each turn, participants chose between two roles. In the \textit{brain} role, the participant selected the piece type to move (e.g., knight or bishop), making the object-level decision, while the AI partner selected the specific move. In the \textit{hand} role, the AI selected the piece type and the participant selected the specific move, making the action-level decision. Across 21 chess players, this design yielded more than 1,100 turn-level decisions in which participants either retained their previous role or switched to the other role. We analyzed gaze behavior, affect, task-specific features, and subsequent task performance around these decisions. We additionally trained a classifier to predict whether participants would switch roles using behavioral and task-specific features.

Our work makes three contributions to human-AI collaboration research. First, we provide an empirical characterization of behavioral and task-level patterns surrounding user-initiated role switches, showing that participants exhibited more exploratory gaze before switching and that switching was associated with lower subsequent move quality. Second, through qualitative interviews, we identify how perceived AI ability, subtask difficulty and importance, and user's desire to perform particular decisions themselves shaped their role choices. Third, we demonstrate that behavioral and task-specific signals can be used to predict whether participants will switch roles and use feature-set ablations to characterize the contribution of different signal categories to predictive performance. Together, these findings suggest that implicit behavioral and task cues can provide useful signals of user's evolving preferences for how work should be divided during sequential human-AI collaborative tasks, informing the design of more user-centered adaptive collaborative systems.

\section{Related Work}
We situate our work at the intersection of role allocation in human--AI collaboration, user preferences for delegation, and behavioral modeling for enabling adaptive collaboration. Prior work across these areas has examined how humans and AI divide work, what factors shape user's preferred division of tasks, and how collaborative systems adapt to changing task conditions. Our work focuses specifically on repeated, user-initiated changes between complementary roles in a sequential decision-making task. 

\subsection{Roles and Dynamic Task Allocation in Human--AI Collaboration}
Humans and AI often possess different capabilities, making them better suited to different subtasks within a collaborative task. Prior work has examined how these differences inform the roles that human and AI teammates take when collaborating on a task~\cite{siemon2022, lindgren2025}. Related work on human-AI complementarity similarly emphasizes that collaboration can benefit when two collaborators contribute distinct information, skills, or capabilities~\cite{hemmer2024}. However, how these complementary subtasks should be distributed between collaborators, particularly when their suitability and desire for involvement in certain subtasks vary as the task progresses, is relatively unexplored. 

Prior work has also explored mechanisms for assigning subtasks between humans and AI or robots based on task characteristics, agent capabilities, or system objectives~\cite{he2023, lamon2023}. Some approaches further support dynamic role allocation, allowing assignments to change as task conditions or collaborator capabilities evolve~\cite{lamon2023}. Mixed-initiative systems similarly seek to make the division of work more flexible by allowing initiative or task allocation to shift during the interaction~\cite{horvitz1999, bradshaw2004, muller2022}. Many such adaptations are determined by the system using signals such as task criticality, confidence, workload, or inferred user ability~\cite{salikutluk_evaluation_2024, hauptman2024}. However, these adaptations may misalign with user preferences, leading to issues with transparency, predictability, and coordination~\cite{tambe2000, hauptman_adapt_2023}.

Our work instead focuses on how users themselves change roles during an ongoing collaboration. Rather then treating roles as fixed or adapting them solely through agent-initiated modifications, we observe repeated, user-initiated transitions between roles that assign different but complementary subtasks to the human and AI. This allows us to characterize when such transitions occur and what behavioral and task-specific signals accompany them.

\subsection{Delegation and Role Preferences}
A growing body of work explores the factors that shape delegation preferences, often through surveys and hypothetical scenarios~\cite{lubars2019, cvetkovic_task_2022, jin2024three, svikhnushina2023expectation}. Trust consistently emerges as a key predictor of delegation behavior, alongside task difficulty, motivation, and perceived risk~\cite{lubars2019, cvetkovic_task_2022}.
Recent longitudinal studies show increasing willingness to delegate as AI capabilities improve. \citet{jin2024three} report that users are more inclined to delegate difficult tasks, especially when trust and motivation are high. Similar patterns are seen in daily interactions with digital assistants~\cite{svikhnushina2023expectation}.
Beyond these factors, cognitive framing also shapes collaboration behavior. \citet{gurney2023role} find that biases such as risk aversion and automation over-reliance affect user effort and performance. These findings underscore the need to study preferences in behaviorally realistic settings.

However, much of this prior work considers delegation at the task level and  captures stated preferences, which may not reflect real-time decision-making in dynamic contexts~\cite{viney2002discrete, de2021stated}. Our study addresses both limitations by observing revealed role preferences in a live, sequential decision-making task in which both the human and AI remain involved while performing complementary subtasks.

\subsection{Modeling Dynamic Human--AI Collaboration}
Adaptive human--AI systems commonly rely on system-level signals such as confidence or workload to determine when adaptation is needed~\cite{amershi2019guidelines, roehr2010using, hauptman2024}. While effective for efficiency, such methods often lack explicit models of user role preferences, and rarely incorporate real-time behavioral signals.

In parallel, user modeling work has focused on predicting trust or intent in interactive systems~\cite{guo2022building, sun2017collaborative, kraus2021modelling}, but these models are generally static and do not address moment-to-moment changes in user's preferred role within a collaborative task. Recent work by \citet{gurney2023role} highlights how framing effects shape trust and reliance, further emphasizing the need for user-centered adaptation.

Building on the four-level delegability framework by \citet{lubars2019}, our setup enables modeling role-switching behavior directly from behavioral and task-specific signals observed during the collaboration. Because participants can repeatedly move between object-level and action-level roles, our setup captures fine-grained changes in how users prefer to allocate work with an AI partner. We examine whether signals such as gaze, affect, and task state characterize these transitions and whether they can be used to anticipate an upcoming role switch. Such signals could support more human-centered adaptive human-AI systems that respond to user's evolving preferences for how work is divided, rather than relying solely on system-level indicators.

\begin{figure*}
    \centering
    \includegraphics[width=0.9\linewidth]{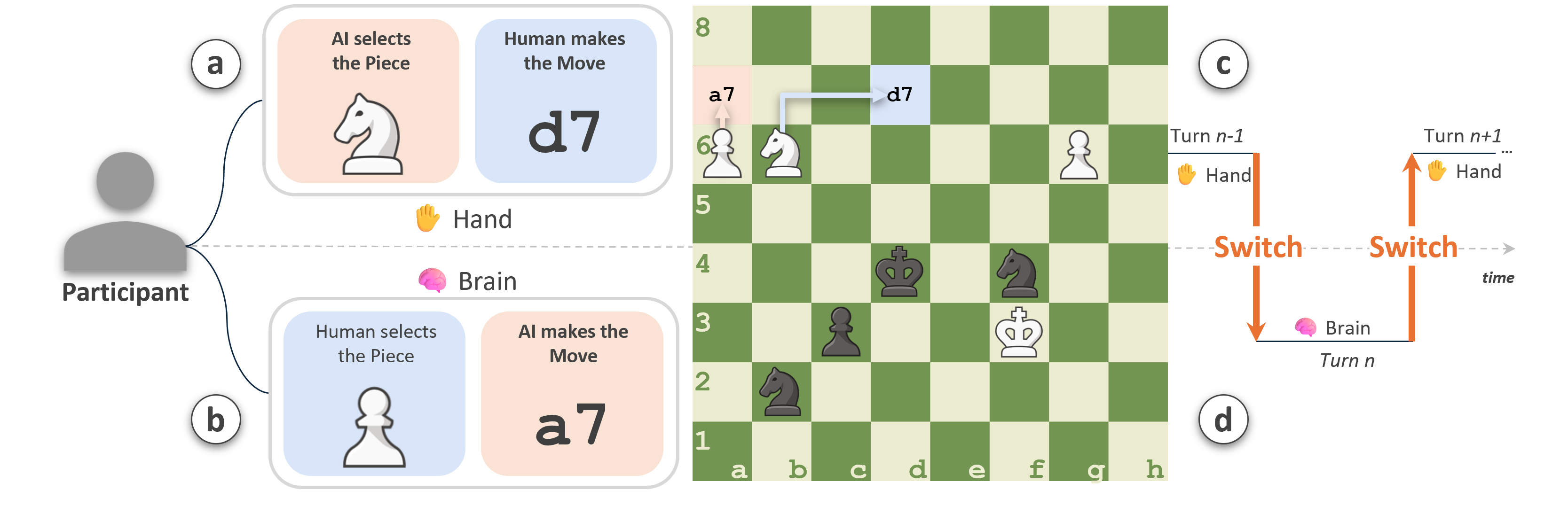}
    \caption{Hand-and-Brain Chess: At each turn, participants chose to act as either the \textit{hand} or the \textit{brain} after evaluating the board. (a) In the \textit{hand} role, the AI selected the piece type, and the participant made a legal move with a piece of that type. (b) In the \textit{brain} role, the participant selected the piece type, and the AI executed a legal move. (c) Participants switched between these two \textit{roles} over time, based on factors that our model uses to predict switching behavior.
}
    \label{fig:task}
    \Description{ The figure illustrates our modified version of hand-and-brain chess. Panels (a) and (b) show examples of the Hand and Brain modes, respectively. The corresponding piece movements are illustrated on the chessboard in panel (d). Panel (c) shows a timeline depicting how participants alternate between the two modes across turns.}
\end{figure*}

\section{Methodology}

We designed a human-subjects study using hand-and-brain chess to examine dynamic role switching during sequential human--AI collaboration. In this this section, we first describe why hand-and-brain chess provides a suitable testbed for studying repeated subtask allocation, then detail how we adapted this variant of chess to elicit sequential role choices, followed by the study materials and procedure.

\subsection{Hand-and-Brain Chess}

Hand-and-brain chess divides each turn into two complementary subtasks, selecting a piece type and making the move with that piece type, performed by two collaborators. In the standard version, these subtasks are allocated before the task begins and remain fixed as the task progresses. We adapted this format so that participants could choose which role to assume at the beginning of every turn. In the \textit{brain} role, the participant made the object-level decision, selecting a piece type (e.g., knight or rook), after which the AI partner executed a legal move with a piece of that type. In the \textit{hand} role, the AI selected the piece type and the participant made the action-level decision by choosing a legal move with that piece type. Thus, both collaborators remained actively involved on every turn, while the subtask assigned to each collaborator depended on the participant's role selection. As shown in Figure~\ref{fig:task}, each turn therefore yielded a decision to either retain the previous role or switch to the other role. This repeated selection enabled us to observe changes in participant's preferred subtask allocation over the course of the game and relate them to evolving behavioral and task-specific context.

\subsection{Materials}

\subsubsection*{Browser Extension}
\begin{figure}
    \centering
    \includegraphics[width=\linewidth]{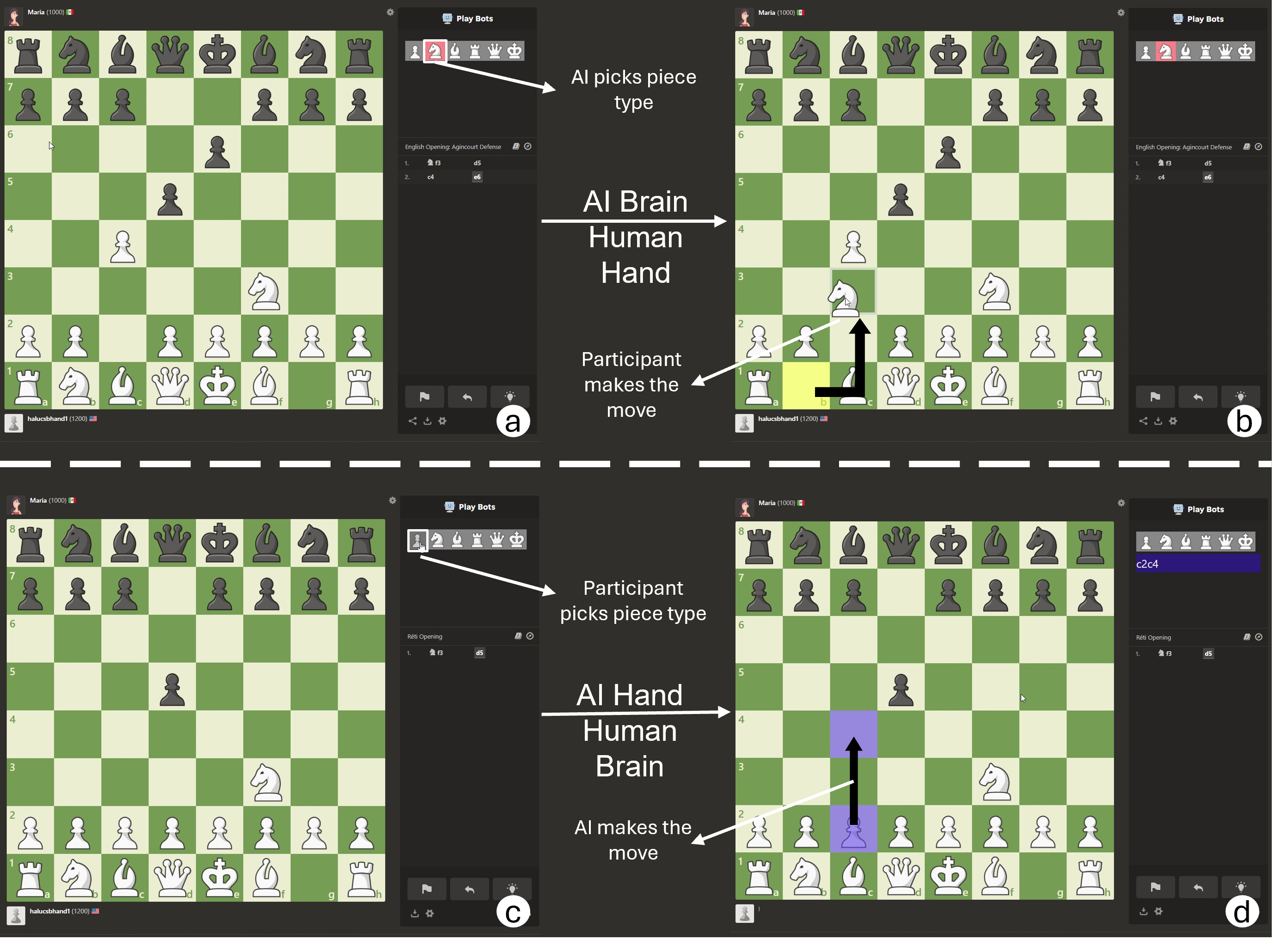}
    \caption{Illustration of human-AI hand-and-brain chess on the Chess.com platform. This figure shows how our browser extension enables hand-and-brain chess for a human participant and an AI teammate. Top: AI brain, participant hand. (a) The AI selects a piece type, highlighted in red, and (b) the participant executes a legal move with that piece type. Bottom: Participant brain, AI hand. (c) The participant selects a piece type (pawn), and (d) the AI’s chosen move is conveyed through highlighted squares on the board and corresponding text below the piece selection area.}
    \label{fig:hand_and_brain_modes}
    \Description{Screenshots from the Chess.com platform demonstrate human–AI hand-and-brain chess played using our custom browser extension. The top row shows the AI selecting a piece type, highlighted in red, and the participant making a legal move with that piece. The bottom row shows the participant selecting a piece type and the AI indicating its move by highlighting the starting and ending squares in blue and displaying corresponding text on the interface.}
\end{figure}

We implemented the experiment using a custom Chrome extension integrated with the Chess.com interface (Figure~\ref{fig:hand_and_brain_modes}). 
The extension recorded all user interactions and cursor positions. The browser extension connected via HTTP to a local chess engine to enable the use of an AI teammate. To ensure adherence with Chess.com's Terms of Use \footnote{https://www.chess.com/legal/user-agreement} and prevent any fair play violations, the extension was restricted to operate only on special research accounts provided by the platform.

\subsubsection*{AI Teammate}

We used the Maia chess engine~\cite{mcilroy2020maia} as the AI teammate for all the participants, playing at a fixed rating of 1500 Elo, which corresponds to an intermediate-level club player. Unlike engines such as Leela Chess Zero~\cite{LeelaChessZero} or Stockfish 17~\cite{Stockfish}, which aim for optimal play, Maia is trained to predict human moves and makes human-like errors, improving the interpretability of its decisions and fostering a more realistic collaborative dynamic. For this reason, participants could not blindly trust the AI, but instead had to consider factors such as the complexity of the position on the board and recent AI behavior before assigning the AI's role.

\subsubsection*{Gaze Tracking}
User gaze was tracked using Beam Eye Tracker\footnote{\url{https://beam.eyeware.tech/get-beam/}}, a software solution that uses webcam input to estimate eye and head movements. We logged gaze position and head orientation at 30 fps via Beam’s Python SDK.

\subsubsection*{Facial Recording}
We recorded participant facial expressions using a webcam positioned above the monitor. These recordings were used to capture any emotional responses during gameplay.

\subsection{Participants}

We recruited twenty-one participants (18 male, 3 female; ages 18–29) from a local chess club. We assessed participant skill level using their blitz rating on Chess.com, as this time control is the most popular on the platform and offers a current and reliable indicator of their skill\footnote{\url{https://www.chess.com/terms/blitz-chess}}. Participant ratings ranged from 400 to 2200 Elo, ranging from novices to advanced club‑level players. Although there is a gender imbalance in participants, it matches the broader gender skew in club-level chess populations~\cite{arnold2024checking}.

\subsection{Procedure} \label{sec:measures}

After providing informed consent (protocol \#anonymous), participants completed a demographic survey and indicated their preferred role (hand/brain) when paired with teammates of lower, similar, or higher rating. We then calibrated eye-tracking using Beam's standard routine to ensure reliable gaze data collection during gameplay.

The study consisted of two phases. In both phases, participants played hand-and-brain chess with the White pieces and partnered with the same AI teammate (Maia~\cite{mcilroy2020maia}) whose strength was fixed at 1500 Elo. Each participant played against a Chess.com bot at a rating as close as possible to their blitz rating. Participants were informed of the rating of the Chess.com bot they would face. However, they were only told that their teammate was an AI but its underlying engine and skill level were not disclosed.
 
In phase 1, participants started by playing four chess positions selected to represent different stages of the game (opening, two middlegame positions, and endgame) with evaluations near $0.0$ by Stockfish 17~\cite{Stockfish}. For each position, participants had one minute to analyze before playing five turns of our modified hand-and-brain chess. This familiarized them with the interface and helped them form an initial mental model of the AI teammate. Data collected during this familiarization phase, including raw gaze positions and role switch decisions, were excluded from subsequent analysis. 

In phase 2, after the four initial familiarization positions, each participant played a full game with the AI teammate, which played at a fixed rating of 1500 Elo. Throughout the full game, we recorded participant facial expressions, eye-gaze, time elapsed, board state, moves, and hand-brain choices.

After the game, participants completed a semi-structured interview with one of the researchers who is a titled player. They were asked to narrate the game, identify key decision moments, and explain their choices to act as the ``brain'' or the ``hand'' at specific moments. To conclude the study, participants completed post-study questionnaires assessing teammate trust and perceived team dynamics. Our post-study surveys combined items from instruments including the Trust in Autonomous Systems (TIAS)~\cite{jian2000foundations} and Cohesion and Satisfaction scales from~\citet{bushe1995appreciative}, with customized questions on strategic alignment. It included 21 Likert-scale items segmented into four sub-scales : alignment (N=4), trust (N=6), cohesion (N=7) and satisfaction (N=4). 


 \section{Data Analysis \& Results}

This section details our data analysis methodology and presents the findings of our study on control level dynamics in human–AI collaboration. 

The experimental design yielded a rich, multimodal dataset. Each game generated approximately 57 decision points with associated gaze patterns, emotional state (derived via post-hoc video analysis of facial expressions), and game state complexity metrics. We analyzed data from all twenty-one players, but excluded one participant (P2, male) from analyses involving gaze metrics due to missing raw gaze data.

\subsection{Pre-study Questionnaires}

Responses from the pre-study questionnaire showed that the participants had limited familiarity with hand-and-brain chess, as most reported playing it infrequently ($\mu = 1.19$ on a 7-point Likert scale, where 1 = rarely and 7 = often). 
Stated preferences for roles in hand-and-brain chess depended on the relative rating of their hypothetical teammates. 13 of the 21 participants stated they would prefer to play as ``brain'' when teamed with a lower-ranked partner (lesser than 100 points) or  a partner with a similar rating (within 100 points). When the hypothetical teammate's rating was higher (more than 100 points), 14 of the 21 participants preferred to play as ``hand''. 

\subsection{Gameplay Statistics}
Of the 21 games played with the AI teammate, participants recorded 13 wins, 4 losses, and 4 draws.  The shortest game lasted 17 moves, while the longest extended to 115 moves. In terms of duration, the quickest game took 3 min and 24s, and the longest lasted 23 min and 52s. In total, participants made 1167 moves, selecting the ``hand'' role for 609 moves and the ``brain'' role for 558 moves. Participants switched roles on 405 moves and retained their previously selected role on 762 moves.


\subsection{Post-study Questionnaires}
\begin{figure}[htbp]
    \centering
    \begin{subfigure}[b]{0.24\linewidth}
        \centering
        \includegraphics[width=\linewidth]{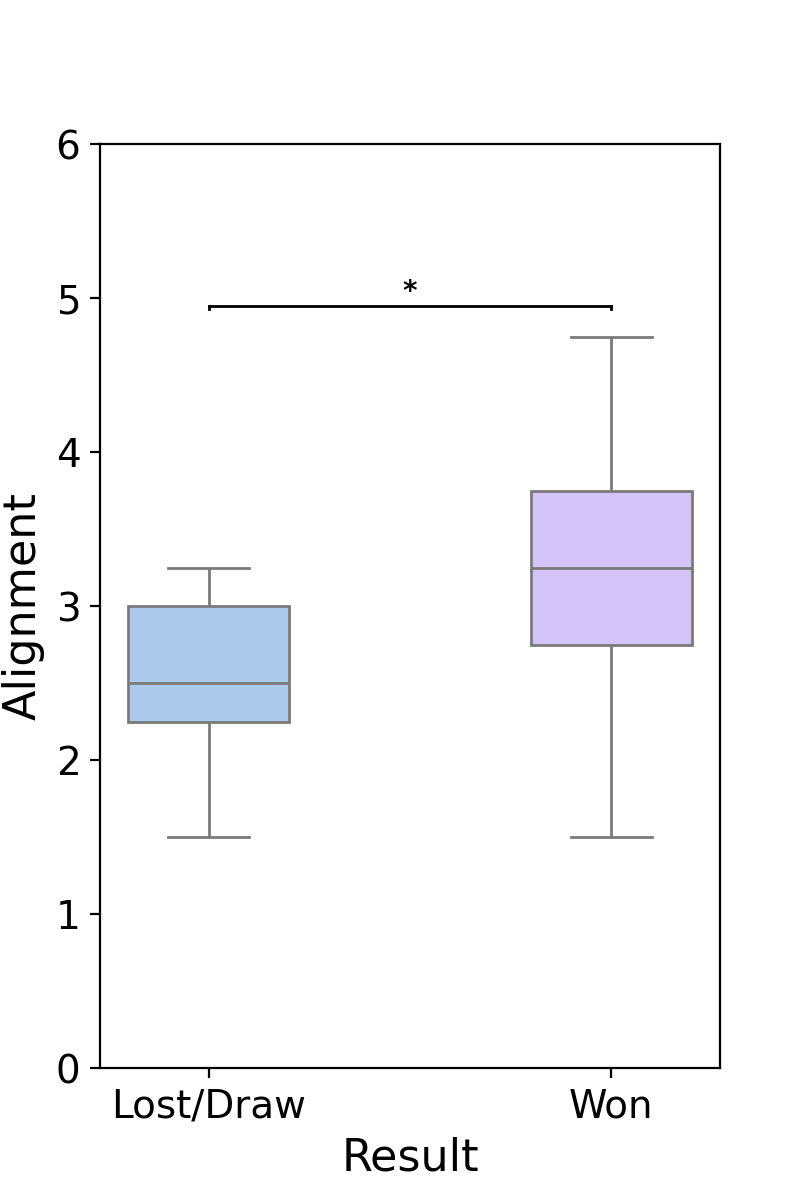}
        \caption{Alignment}
        \label{fig:alignment}
    \end{subfigure}
    \hfill
    \begin{subfigure}[b]{0.24\linewidth}
        \centering
        \includegraphics[trim=0 10 0 10, width=\linewidth]{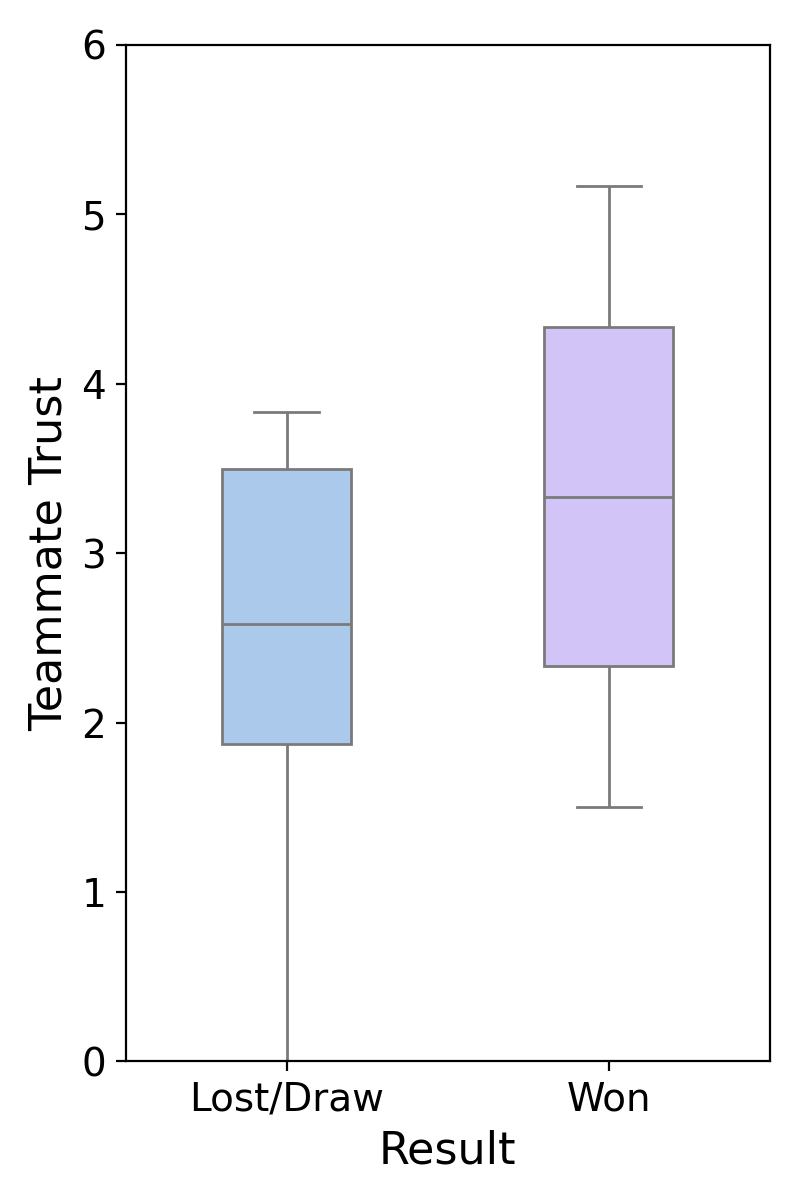}
        \caption{Trust}
        \label{fig:trust}
    \end{subfigure}
    \hfill
    \begin{subfigure}[b]{0.24\linewidth}
        \centering
        \includegraphics[width=\linewidth]{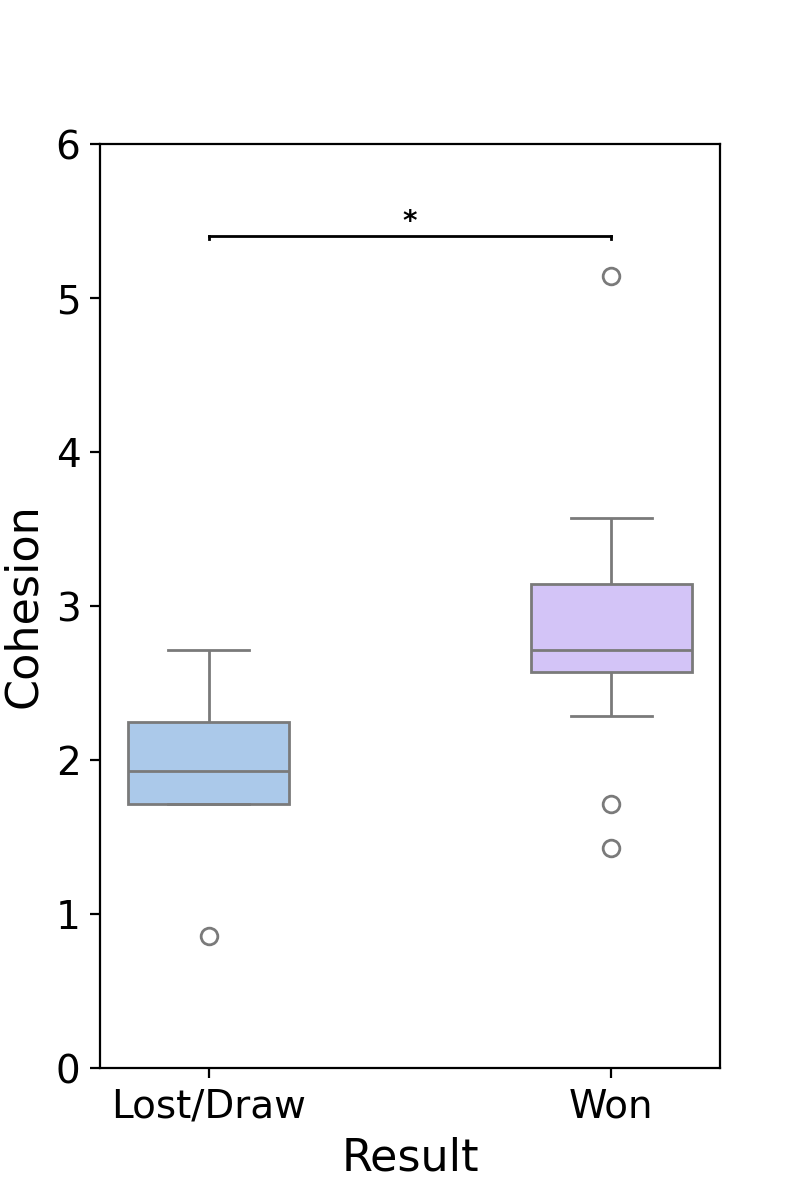}
        \caption{Cohesion}
        \label{fig:cohesion}
    \end{subfigure}
    \hfill
    \begin{subfigure}[b]{0.24\linewidth}
        \centering
        \includegraphics[width=\linewidth]{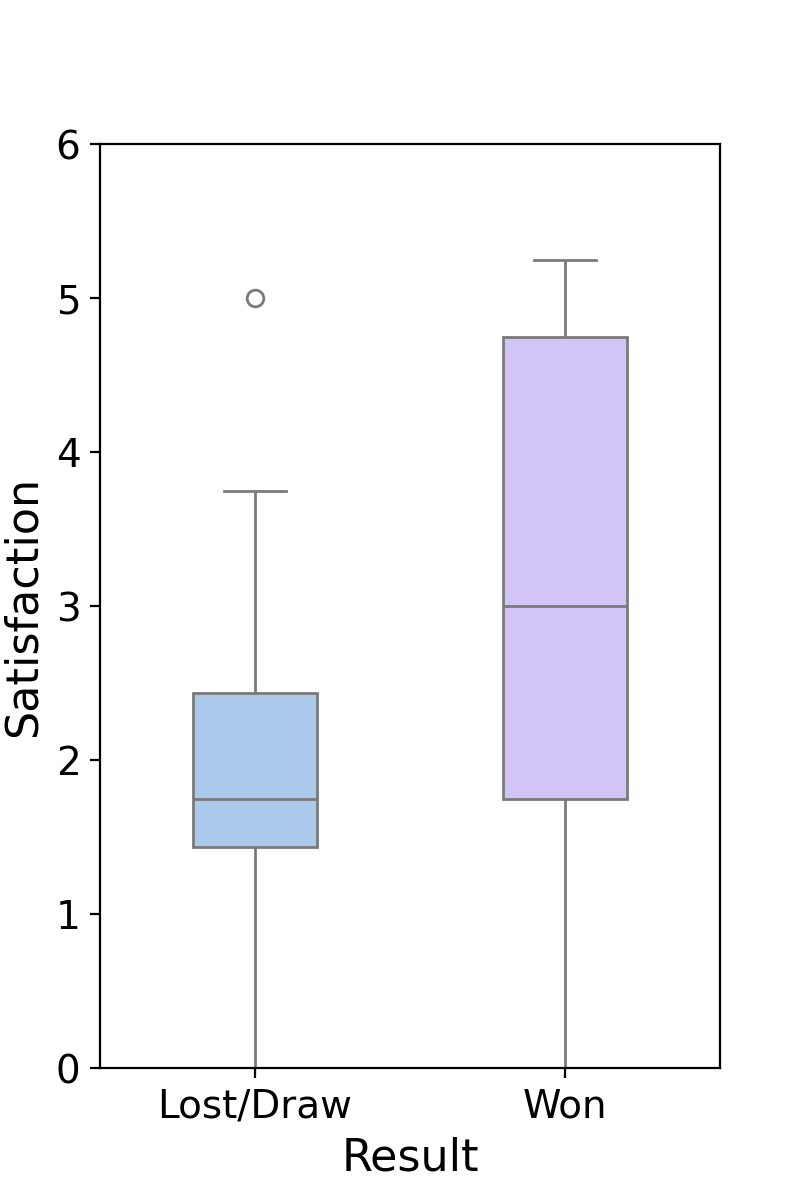}
        \caption{Satisfaction}
        \label{fig:satisfaction}
    \end{subfigure}
    \caption{Post-study questionnaire ratings for alignment, trust, cohesion, and satisfaction. Ratings are shown by game outcome (win vs.\ loss/draw), with higher scores indicating more positive perceptions of the AI teammate.}
    \label{fig:post_study}
\end{figure}

To evaluate the subjective quality of the participant-AI partnership, we used post-study questionnaires (Section~\ref{sec:measures}) to assess teammate alignment, trust, cohesion, and satisfaction. Each factor was measured using multi-item, 7-point Likert scales which were linearly rescaled to a 0–6 range for analysis. A composite score (out of 6) was computed for each sub-scale by averaging the individual item scores. Since a participant's experience of their AI teammate may be shaped by the result, we additionally ran a linear regression for each sub-scale with the game outcome (won vs lost/drew) as predictor. This allows us to assess whether these subjective ratings reflect a genuine perception of the partnership or whether they are significantly impacted the game outcome. The results are shown in Figure~\ref{fig:post_study}

\paragraph{Alignment} Participants rated the overall alignment with their AI teammate's moves as neutral ($\mu = 3; \sigma = 0.89$). However, perceived alignment appeared to be influenced by the game's outcome. Participants who won their game reported higher alignment scores ($\mu=3.29$) compared to those who did not ($\mu=2.53$). A linear regression confirmed this trend, with winning associated with $0.76$-point increase in alignment score (($b=0.76$, $SE=0.37$, $t=2.03$, $p=.056$, 95\% CI$[-0.02, 1.54]$), though the effect was not significant. 

\paragraph{Trust} The overall trust score ($\mu = 3; \sigma = 1.3$) indicates a lack of trust in the AI teammate across participants. Participants who won their game, on average, trusted their teammate more ($3.36$) than those who did not win ($2.44$ ). A linear regression showed that winning was associated with a $0.92$-point increase in trust score ($b=0.92$, $SE=0.56$, $t=1.64$, $p=.117$, 95\% CI$[-0.25, 2.10]$) but this effect was not statistically significant. 

\paragraph{Cohesion} The overall team cohesion score was low ($\mu = 2.4; \sigma = 0.9$), with participants who won their game reporting higher scores ($\mu = 2.82$) than those who did not ($\mu = 1.94$). This difference was reflected in linear regression, where winning was associated with a $0.88$-point increase in cohesion score ($b=0.88$, $SE=0.36$, $t=2.44$, $p=.025$, 95\% CI$[0.12, 1.63]$), a statistically significant effect.  

\paragraph{Satisfaction} The overall satisfaction score was low ($\mu = 2.72; \sigma = 1.73$), indicating that all the participants were dissatisfied with their AI teammate's performance. A linear regression indicated that winning was associated with a $0.97$-point increase in satisfaction score ($b=0.97$, $SE=0.77$, $t=1.27$, $p=.220$, 95\% CI$[-0.63, 2.57]$), though the difference was not statistically significant.

\subsection{Quantitative Analysis}\label{sec:quant_analysis}

To construct a dataset for statistical analysis, we first time-synchronized the facial recordings, raw gaze data, and participant interaction logs, including moves and collaboration mode selections. We then estimated the participant's emotional state using Deep Face~\cite{serengil2024lightface} on every \nth{10} frame of the facial video recording. These sampled frames were pooled and temporally aligned with each role selection decision and turn, ensuring correspondence between emotion estimates and decision intervals. DeepFace~\cite{serengil2024lightface} outputs a probability distribution over seven emotion categories. From this distribution, we extracted the probability of surprise, after the opponent's turn, as it was the most relevant for estimating the participant's response to the turn. Surprise is often defined as a reaction to disconfirmed expectations, when an outcome deviates from what is likely or expected \cite{meyer1997toward, teigen2003surprises, gross2014impact}. However, other research suggests that surprise may also reflect a breakdown in sense-making, where the level of surprise corresponds to the difficulty of integrating new information into an existing mental model~\cite{maguire2011making}. In the context of our study, both perspectives are relevant. Unexpected opponent decisions may have surprised participants either because they defy prior expectations or because they disrupt ongoing strategy formulation. 

We processed the dataset to extract data associated with the turns that required participants to consider several possible moves from multiple pieces. We excluded two types of moves, the first move of the game, as it lacks a prior context, and moves made in a decisively won or lost position. Such positions were identified by either a significant material imbalance (e.g., King and Queen vs King) or extreme evaluations (e.g., +10, which indicates that the white side has an advantage equivalent to an extra queen and pawn) from Stockfish 17, where the mode choice would be inconsequential and would not change the outcome. This filtering removed 154 turns, resulting in a final dataset of 992 role selections since players need to select a role each turn (524 hand and 468 brain selections; 341 switches, 651 non-switches) for analysis. 

Using this dataset, we examined whether individual behavioral and task-specific metrics differed between switch and non-switch turns. Depending on the expected metric distribution and model fit diagonsitics, we used linear mixed-effects models (LMMs) or generalized linear mixed-effects models (GLMMs), with participant included as a random intercept. We used mixed-effects models because the move-level observations were nested within participants and therefore were not independent. These analyses characterize differences in behavioral and task-state measures immediately before or after participant's role-selection decisions.

\begin{figure}[htbp]
    \centering
    \begin{subfigure}[b]{0.3\linewidth}
        \centering
        \includegraphics[width=\linewidth]{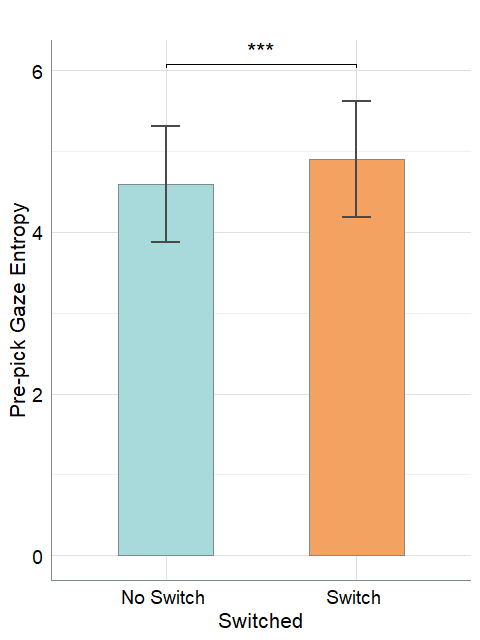}
        \caption{Pre-pick gaze entropy}
        \label{fig:entropy}
    \end{subfigure}
    \hfill
    \begin{subfigure}[b]{0.3\linewidth}
        \centering
        \includegraphics[width=\linewidth]{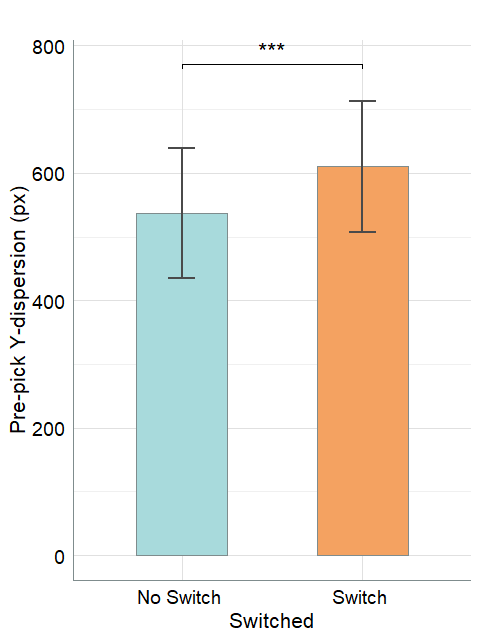}
        \caption{Y-Dispersion}
        \label{fig:disp}
    \end{subfigure}
    \hfill
    \begin{subfigure}[b]{0.3\linewidth}
        \centering
        \includegraphics[width=\linewidth]{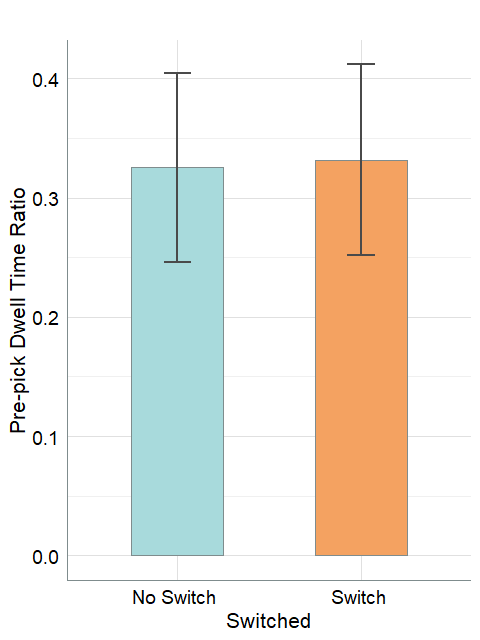}
        \caption{Dwell Time Ratio}
        \label{fig:dwell}
    \end{subfigure}
    
    \caption{Gaze Metrics : Pre-pick gaze entropy and dispersion along the vertical axis were significantly different between switch and non-switch turns}
    \label{fig:gaze}
\end{figure}

\subsubsection*{Gaze Patterns before a Role Switch}
In prior work, gaze behavior has been used to infer attention allocation~\cite{rayner1998eye,just1980theory} and is also recognized as a marker of cognitive state, including task difficulty and decision-making effort~\cite{marshall2007identifying,goldberg1999computer}. Therefore, in the context of human-AI collaboration, we hypothesized that gaze patterns would differ before participants made a decision to switched roles compared to when they did not, reflecting increased deliberation or cognitive conflict preceding a mode change. 
We analyzed differences in pre role-selection gaze patterns between switch and non-switch selections using LMMs, with participant included as a random intercept to account for repeated observations within participants. Model fit was assessed by visually inspecting the Q-Q plots of the residuals. Pre-selection gaze entropy was significantly higher ($\chi^2(1) = 20.47, p < 0.001$) when participants switched roles ($M = 5.04, SD = 1.61$) than when they did not ($M = 4.78, SD=1.58$). Similarly, the gaze dispersion along the vertical length of the board was significantly higher ($\chi^2(1) = 17.59, p < 0.001$) before switching ($M = 632.98$px, $SD=310.02$px) than before non-switches ($M = 563.06$px, $SD=313.5$px), confirming our hypothesis. However, the ratio of gaze dwell time to thinking time (the duration from the opponent's last turn to the participant picking a role) was not significantly different in turns when users switched roles compared to turns when they did not (Switch $= 0.35 \pm 0.21$, No Switch $= 0.33 \pm 0.22$ ; $\chi^2(1) = 0.36, p=0.55$). The results are shown in Figure~\ref{fig:gaze}.

\subsubsection*{Position Complexity and Role Switching}
We used chess fragility score~\cite{barthelemy2025fragility} to quantify the complexity and critical nature of the positions that occurred during each chess game. Fragility score ranges from 0 to 1, with higher values indicating more decisive positions in a game. It is a robust metric that remains consistent across games with different openings and players of various skill levels~\cite{barthelemy2025fragility}. We hypothesized that participants would be more likely to switch roles in more fragile positions. We examined fragility both immediately before role selection and after the subsequent move. As LMMs for both measures exhibited non-normal residuals based on visual inspection of Q-Q plots, we instead used Tweedie GLMMs with a log link and participant-level random intercepts. Pre-selection position fragility did not significantly differ between switch and non-switch turns ($\beta = 0.176$, $SE = 0.115$, $z = 1.53$, $p = .126$), providing no evidence that greater position fragility was associated with participant's tendency to switch roles. In contrast, post-move fragility was significantly higher on switch turns ($\beta = 0.269$, $SE = 0.116$, $z = 2.32$, $p = .021$). Given the log link, exponentiating this coefficient ($e^{0.269}=1.308$) indicates that expected post-move fragility was approximately $30.8\%$ higher following a role switch than following a non-switch turn.

\subsubsection*{Impact of Switching on Position Evaluation}

To examine the effect of role switching on subsequent turn quality, we analyzed changes in position evaluation evaluated by the Stockfish 17 chess engine. In addition to the earlier exclusion of decisive positions, we excluded turns with extreme changes in evaluation resulting from Stockfish evaluations approaching $\pm29,000$ centipawns. These values produced disproportionately large $\Delta \mathrm{eval}$ values that hindered reliable model estimation. Raw evaluation differences were first converted into centipawn loss (CPL)\footnote{\url{https://www.chess.com/blog/raync910/average-centipawn-loss-chess-acpl}} by bounding improvements at zero ($\mathrm{CPL} = \max(0, -\Delta\mathrm{eval})$), ensuring non-negative values where zero represents optimal play and higher values represent larger errors. This transformation produces non-negative continuous values concentrated around zero which corresponds to optimal moves. We therefore modeled CPL using a Tweedie GLMM with a log link function. 

Descriptively, the CPL was higher on turns in which users switched roles ($M = 78.9, SD=117.38$) than on non-switch turns ($M = 59.93, SD=96.1$). The model revealed a significant effect of role switching on CPL ($\beta = 0.253, SE=0.091, z=2.79, p=0.005$). Given the log link, exponentiating the coefficient indicates that switching roles was associated with a $28.8\%$ increase in expected CPL relative to non-switch turns. These results suggest that, within participants, role switching was associated with lower subsequent move quality.

\subsubsection*{Surprisal and Role Switching}
To investigate the role of surprise in mode switching, we computed the average surprise probability in the time window between the end of the opponent's turn and the participant's role selection. Surprise did not significantly differ between switch and non-switch turns ($\beta = 0.018$, $SE = 0.13$, $z = 0.14$, $p = .88$). Thus, we found no significant difference in estimated surprise immediately preceding role selection between switch and non-switch turns.

\subsection{Thematic Analysis of Interview Data}

Because quantitative measures may not fully capture the influence of implicit biases, trust, and other subjective factors on role selection, we analyzed the semi-structured interview recordings using reflexive thematic analysis~\cite{braun2006using}. This approach allowed us to identify recurring patterns in how participants formed and revised their role preferences over the course of the task.

Following guidelines by~\citet{braun2006using}, two researchers independently reviewed the interview recordings and extracted statements related to participant's role-selection rationales and changes in role preference. These statements were labeled with ``when'' and ``why'' codes to capture when participants preferred particular roles and why. The researchers then independently clustered the codes into candidate themes, compared and reconciled their interpretations, and produced a unified set of three themes.

\paragraph{\textbf{T1: Cognitive Effort and Perceived Decision Risk.}}

Participant role choices were often influenced by the cognitive effort and perceived risk associated with a position. Participants frequently selected \textit{hand} role when faced with many options, during openings, or when unsure what to do, particularly when several legal moves were available across piece types. P7 described this explicitly: ``there are moments where I'm not sure what... there's too many choices, I'm like why don't you make a choice for me so I can just narrow down which piece of that type to move.'' Similarly, P3 preferred the \textit{hand} role in ``a more open position where there were more options,'' while P8 selected \textit{hand} throughout the opening, reasoning that ``there could be so many variations in the opening... you give me whatever you want me to and I'll just adjust to it.'' For P2, this reduction in choice was particularly valuable under time pressure because the \textit{hand} role ``narrowed it down... so I didn't have to think, because time is probably my greatest weakness.''

In contrast, \textit{brain} role was picked by the participants in situations where they wanted to retain control over which piece moved such as defending a valuable piece or trying to avoid a blunder. P9 selected \textit{brain} ``so that I can limit the amount of moves that the AI could screw up in,'' concerned that in \textit{hand} mode the AI would select a piece that would lead to a significant disadvantage. P19 applied the same logic proactively in the middlegame, switching almost entirely to \textit{brain} and ``choosing the plays that... the computer [as] the hand couldn't really lose a good piece." P11 selected \textit{brain} in positions where the number of legal moves was low, such as when in check, where ``there's only two pieces I can use except moving the king.''

These accounts suggest that cognitive effort and perceived risk shaped participant's role preferences over the course of the task. The \textit{hand} role was often preferred when participants want to reduced their own decision space, whereas the \textit{brain} role was preferred when they wanted to obtain control over the piece-selection decision and constrain the AI's subsequent options in consequential positions. 

\paragraph{\textbf{T2: Perceived AI Competence and Intent.}} 
Participant role choices also depended on their evolving perceptions of the AI partner's competence. Several entered the task assuming the AI was at least competent, and in some cases stronger than themselves. P20 recalled that ``at the beginning, honestly, I thought it was maybe a little bit better,'' while P21  reasoned that ``since my elo rating is low, I thought that the bot would be higher than me.'' Others developed an initial level of trust from AI's early play, describing it in terms usually reserved for a strong player. P5 noted that ``it didn't make any blunders as far as I could tell'' and appeared ``definitely above beginner strength,'' while P15 described simply ``trusting the engine for a while'' and assuming its decisions reflected an underlying plan.

These expectations were continually revised through interaction. For some participants, a small number of unexpected moves substantially changed their role selection logic. P9 traced their entire shift toward constraining the AI's available options to an early move that ``lost me my knight.'' P1 similarly came to view the AI teammate as increasingly unreliable, describing it as a ``guy [who] sucks,'' and perceiving its moves as becoming ``worse and worse''. P6 characterized the AI's behavior as ``either reasonable or really, really stupid,''. Both  P1 and P6 responded by shifting into a defensive strategy, trying, in P6's words, ``to force the AI to do what I want it to do.''

For a subset of participants, difficulty explaining the AI's inconsistent behavior led them to attribute intentionality to its poor moves rather than simply judging it as less competent. P1 became convinced that the AI was deliberately making worse moves as the game progressed, while P6 interepreted the contrast between apparently strong and inexplicably poor decisions as evidence that the AI could make good decisions but was sometimes choosing not to. Others entertained this possibility more tentatively. P4, for example, described a particular hypothetical move as one that ``would definitely sabotage me," without concluding that the AI was intentionally doing so. These accounts suggest that when participants could not readily reconcile the AI's behavior with their expectations of its competence, some constructed explanations about the AI's intent that further shaped how cautiously they allocated subtasks.  

Other participants responded differently to the uncertainty in AI's competence. P7 acknowledged thaat ``really can't tell how good they are, because I don't have time to really think through the move they pick'' but treated this uncertainty as a reason to continue relying on the partnership reasoning that the ``AI might see something longer than I did''. P8 described a positive trajectory, becoming increasingly confident as the game progressed because the AI's decisions aligned closely enough with their own intent to make the interaction feel ``streamlined.'' These contrasting trajectories suggest that role preferences were shaped not only by prior expectations of AI competence, but also by how those expectations were updated through interaction. 

\paragraph{\textbf{T3: Probing, Communication Constraints, and Situational Pressure.}} 

Beyond immediate considerations of decision risk and perceived AI competence, participant's role choices were also shaped by longer-term strategies and situational pressures within the game. Some participants deliberately chose particular roles on turns they considered to have obvious moves in order to probe the AI's reasoning or competence, using the resulting behavior to inform later role selections. P5, for instance, chose \textit{hand} specifically ``to see if it knows to move [a hanging piece] away.'' 

Because the hand-and-brain format limited communication to piece-type selection rather than an intended square or plan, participants also had to infer the AI's intent from its choices. P10 wondered, ``why does it want me to move that piece? ... maybe there's some deeper reasoning I didn't think about,'' while P11 felt that the AI ``wasn't kind of understanding me,'' and P15 characterized an unexpected move as a ``nuisance''. For some participants, this ambiguity also affected how comfortable participants felt in their chosen role. P20 reported that \textit{hand} role increased decision anxiety because once AI selected the piece ``then you start second-guessing a little bit''. They instead preferred the \textit{brain} role as it allowed them to "have some idea in your head and watch it probably get executed." 

Participant's role selection strategies were frequently overridden by situational pressure. P13 described the \textit{brain} role as requiring ``double thinking'' since it meant both selecting a piece and modeling what the AI might then do with it, and found that time crunches pushed them back toward \textit{hand} simply to avoid that overhead. P16 responded to time pressure differently, using \textit{brain} to speed up play and trusting the AI to make a move quickly ``even if it made the not-so-best decision, it would be a faster decision.'' 

\subsection{Role Switching Prediction}

From the dataset used for the quantitative analysis, we excluded data from participant P2 (male) due to missing gaze data. This resulted in a final dataset of 971 role selections (341 switches) from the remaining 20 games. 

\subsubsection*{Feature Engineering}

We constructed multimodal features representing participant's recent behavioral history, their behavior during the current turn, and the current task state. Contextual features summarized information over the preceding $k$ turns ($k \in \{3,5\}$), including changes in gaze dispersion, fixation behavior, position evaluation, fragility, and prior role-selection history. Current-turn features were computed over the thinking window between the opponent's preceding move and the participant's role-selection decision and included gaze entropy, gaze dispersion, time spent thinking, remaining clock time, and current position evaluation.

To represent the chess position without manually enumerating a large set of domain-specific board features, we additionally extracted 1024-dimensional learned embeddings of the current position from the penultimate layer normalization of the Maia 2~\cite{tang2024maia} Rapid model. These embeddings provide a learned representation of the current board state. We retained fragility as an explicit feature because it provides an interpretable measure of position criticality that is not explicitly represented as such by the Maia embedding.

To approximate real-time deployment, we generated prediction samples at one-second intervals during each turn. For a turn lasting $T$ seconds, samples were generated at $t=1,\ldots,T$, with each sample containing only information available up to that point in the thinking window. All samples from a turn were assigned the eventual binary outcome indicating whether the participant switched roles relative to the previous turn. Elapsed thinking time was also included as a feature, allowing the model to capture changes in predictive signals over the course of the thinking period. This procedure resulted in 5,935 samples, of which 39.2\% corresponded to switch events.

For subsequent modeling and ablation analyses, we organized the extracted features into five feature sets. \textit{Maia} comprised the 1024-dimensional learned board-state representation. \textit{Gaze} captured eye-movement dynamics during the thinking window, including gaze entropy, fixation and dwell characteristics, and spatial dispersion. \textit{AOI} summarized the spatial distribution of visual attention across predefined interface regions, including the chessboard, interaction log, and regions outside these areas. \textit{Time} captured temporal aspects of thinking, including elapsed thinking time and remaining clock time. Finally, \textit{Chess Context} represented explicit task-state and interaction-history information, including engine evaluation, position fragility, participant rating, rating difference, and prior role-selection behavior such as switch, brain, and hand counts.

\subsubsection*{Model}

We trained a CatBoost gradient-boosted decision-tree classifier~\cite{prokhorenkova2018catboost} because the feature representation combined heterogeneous behavioral, temporal, and learned task-state features and could contain nonlinear relationships and interactions among predictors. To address class imbalance and emphasize difficult switch examples, we optimized the model using a time-weighted focal-loss objective~\cite{lin2017focal},

\begin{align}
\mathcal{L}_{\mathrm{focal}}(p_t,y)
    &= -\alpha_y(1-p_{t,y})^\gamma
       \log(p_{t,y})\,w(t), \\
w(t)
    &= \exp\left(s\frac{t}{t_{\max}}\right),
\end{align}

where $p_{t,y}$ denotes the predicted probability assigned to the ground-truth class at time $t$. The weighting function $w(t)$ assigns greater weight to observations later in the thinking period when $s>0$. The class-weighting parameter $\alpha$, focusing parameter $\gamma$, and temporal-weighting parameter $s$ were tuned during cross-validation.

Because the raw feature representation was high dimensional, particularly due to the 1024-dimensional Maia embeddings, we performed tree-based feature selection using only the training data within each outer cross-validation fold. Features were ranked using CatBoost feature importance from a preliminary model (depth $=4$, 100 boosting iterations, learning rate $=0.05$). We evaluated feature-set sizes between $K=25$ and $K=35$ and retained the top $K=30$ features for the final model. The selected features included dimensions of the Maia representation alongside behavioral and interaction-history features, including prior switching behavior and gaze-based spatial-attention measures.

\subsubsection*{Cross-Validation and Hyperparameter Optimization}

We evaluated model performance using nested, participant-disjoint cross-validation to prevent leakage between repeated observations from the same participant. The outer evaluation used five-fold \texttt{StratifiedGroupKFold} cross-validation, with participant ID as the grouping variable and switch status as the stratification target. Thus, all moves and all one-second samples from a participant appeared entirely within either the training or test partition of a given fold. The folds approximately preserved the overall switch prevalence of $39.2\%$.

Within each outer training fold, model hyperparameters, feature selection, and decision-threshold calibration were performed using an inner three-fold \texttt{StratifiedGroupKFold}. We conducted 20 iterations of randomized hyperparameter search over tree depth, learning rate, number of boosting iterations, $L_2$ regularization, random strength, and focal-loss parameters. Because switch samples constituted the minority class, we used area under the precision--recall curve (AU-PRC) as the primary metric for hyperparameter selection. Classification thresholds were calibrated using precision--recall performance within the inner cross-validation procedure, such that the held-out outer test participants were not used for either hyperparameter or threshold selection. The optimized hyperparameters are listed in Table~\ref{tab:catboost_hyperparams}.

\begin{table}[t]
\centering
\caption{Optimized CatBoost and focal-loss hyperparameters selected through nested participant-disjoint cross-validation.}
\label{tab:catboost_hyperparams}
\begin{tabular}{lc}
\toprule
\textbf{Parameter} & \textbf{Selected Value} \\
\midrule
Learning rate & 0.08 \\
Tree depth & 5 \\
Boosting iterations & 80 \\
$L_2$ regularization & 1.0 \\
Random strength & 0.1 \\
Focal-loss $\alpha$ & 0.10 \\
Focal-loss $\gamma$ & 4.0 \\
Time-weight scale $s$ & 2.0 \\
\bottomrule
\end{tabular}
\end{table}

\subsubsection*{Model Evaluation}

We report AU-PRC, AUROC, precision, recall, F1-score, and balanced accuracy. Because switch samples constituted $39.22\%$ of the dataset, the expected AU-PRC of a random-ranking classifier is $0.39$, while chance AUROC is $0.50$. A majority-class classifier that always predicts \textit{no switch} would achieve $60.78\%$ accuracy but $0\%$ recall for switch events and a balanced accuracy of $0.50$.

Across the five participant-disjoint test folds, the complete model achieved a mean AU-PRC of $0.56 \pm 0.06$ and a mean AUROC of $0.66 \pm 0.02$, exceeding the corresponding random-ranking baselines. At the calibrated classification thresholds, the model achieved $50.35\%$ precision, $55.8\%$ recall, an F1-score of $0.53$, and balanced accuracy of $60.14\%$.

\subsubsection*{Feature-Set Ablation}

To examine the contribution of different information sources, we evaluated five feature-set configurations using the same participant-disjoint five-fold cross-validation procedure. The complete model combined all five feature sets defined above: \textit{Maia}, \textit{Gaze}, \textit{AOI}, \textit{Time}, and \textit{Chess Context}. We compared this model against four reduced configurations: (1) \textit{Maia} alone, (2) \textit{Gaze + AOI + Time}, (3) \textit{Gaze + AOI + Time + Chess Context}, and (4) \textit{Time + Chess Context}. The results are shown in Table~\ref{tab:ablation_results}.

The Maia embeddings accounted for much of the predictive performance, while incorporating behavioral and explicit task-context features provided a modest additional improvement. The complete model achieved a mean PR-AUC of $0.56$, compared with $0.51$ for the Maia-only model, suggesting that behavioral and contextual signals provided complementary predictive information beyond the learned board-state representation.

\begin{table}[t]
\centering
\caption{Five-fold participant-disjoint ablation results. Values are mean $\pm$ SD across folds. Higher is better.}
\label{tab:ablation_results}
\resizebox{\columnwidth}{!}{
\begin{tabular}{lcccccc}
\toprule
\textbf{Feature Set} & \textbf{PR-AUC} & \textbf{ROC-AUC} & \textbf{Precision} & \textbf{Recall} & \textbf{F1} & \textbf{Bal. Acc.} \\
\midrule
Maia + Gaze + AOI + Time + Chess
& $\mathbf{0.56 \pm 0.06}$ & $\mathbf{0.66 \pm 0.02}$ & 50.35\% & 55.80\% & $\mathbf{0.53}$ & $\mathbf{60.14\%}$ \\

Maia only
& $0.51 \pm 0.04$ & $0.61 \pm 0.03$ & 50.78\% & 37.76\% & 0.43 & 57.07\% \\

Gaze + AOI + Time
& $0.42 \pm 0.08$ & $0.51 \pm 0.03$ & 45.36\% & 34.45\% & 0.392 & 53.83\% \\

Gaze + AOI + Time + Chess
& $0.46 \pm 0.08$ & $0.57 \pm 0.04$ & 44.82\% & $\mathbf{58.38\%}$ & 0.507 & 56.00\% \\

Time + Chess
& $0.43 \pm 0.09$ & $0.55 \pm 0.05$ & 42.29\% & 25.09\% & 0.315 & 51.49\% \\
\bottomrule
\end{tabular}
}
\end{table}

\section{Discussion}

Our findings suggest that role switching in sequential human--AI collaboration emerges from user's evolving preferences for how work should be divided across subtasks. In our task, the \textit{brain} role corresponded to the object-level decision of selecting which piece type to move, whereas the \textit{hand} role corresponded to the action-level decision of selecting the specific move. The qualitative analysis primarily captured how participants formed and revised these preferences over the course of the game, while the quantitative analyses examined the behavioral and task-level conditions associated with observed switches. Taken together, the results point to four broader implications for the design of human-centered mixed-initiative systems.

\subsection{Preferred roles depend on subjective difficulty more than objective task state }

Participants frequently described cognitive effort, uncertainty, and perceived decision risk as factors shaping their role choice. The action-level role (\textit{hand}) was often preferred when participants wanted the AI to narrow their decision space, whereas the object-level role (\textit{brain}) was preferred when they wanted to retain control over a consequential higher-level decision or constrain the AI's options. 

However, these qualitative accounts were not mirrored by our objective measure of position complexity. Pre-role-selection fragility of the position did not significantly differ between switch and non-switch turns. This divergence suggests that objective task complexity and perceived difficulty are not necessarily equivalent. Prior work has similarly shown that user's subjective assessments of task difficulty can diverge from object task state and can affect how they allocate work to AI~\cite{fugener2019, song2026}.
During the game, two positions with comparable fragility may have imposed very different cognitive demands depending on the participant's experience, current strategy, confidence, or their perception of AI's competence. 

The gaze findings offer a complementary view. Participants exhibited higher gaze entropy and greater vertical gaze dispersion before switching, while the dispersion was the only feature that made a significant contribution in the multivariate switching model. This is consistent with broader visual exploration during moves when participants were reconsidering their role~\cite{shiferaw_review_2019, brunye_eye_2017}. 

Our results suggest that adaptive human--AI collaborative systems should not infer user's preferred division of work solely from objective task complexity measures. Combining task-state information with user-specific behavioral signals may provide a better indication of how user's preferred roles are changing. More broadly, systems should support flexible allocation at the subtask level rather than treating delegation as an all-or-nothing decision that remains static throughout the task. Decomposing tasks into complementary higher-level and action-level components can give users more expressive ways to retain the parts of a decision they consider consequential while delegating others.
\subsection{Mental model formation via probing}
Several participants began the task with assumptions about the AI's relative ability but revised these judgments as they observed its behavior. Unexpected or poor decisions led some participants to select object-level subtask in order to constrain the AI's available options, while others continued relying on the AI because they interpreted unexpected behavior as potentially reflecting insight or skill they themselves lacked. Some users opted to probe the AI's competence by choosing specific roles on moves with seemingly obvious moves, hoping to use the AI response to inform later selections. 

This reflects a broader challenge in human--AI collaboration where users often begin an interaction without a calibrated understanding of where an AI system performs well, where it is unreliable, or how its capabilities vary across subtasks. This requires users to construct and continuously revise their mental model of the AI's capabilities through prior experience and repeated observation~\cite{bodamer2026}.   

The post-study questionnaire results provide additional context for these evolving judgments. Participants reported moderate alignment and trust overall, but relatively low cohesion and satisfaction. Ratings were consistently higher among participants who won their game, although game outcome was significantly associated only with cohesion. These results suggest that perceptions of the AI partner were shaped by the interaction as a whole rather than by isolated AI actions alone. This pattern is consistent with prior human--AI research which have identified communication, coordination and shared cognition as persistent challenges~\cite{schmutz2024, mcneese_whowhat_2021}.

Participant's role choices likely reflected a trade-off between exploiting their current understanding of the AI and exploring its capabilities. In some cases, participants chose roles that constrained the AI based on prior experience; in others, they deliberately selected roles that allowed them to observe how the AI would behave and update their expectations. This suggests that role allocation can serve not only to divide work, but also to learn about an AI collaborator. Collaborative systems could therefore support low-consequence opportunities for users to probe different role allocations and learn about the AI's capabilities before relying on those allocations in more consequential situations.
\subsection{Role switching can impact performance}
Switch turns were associated with higher subsequent centipawn loss indicating that changes in role were followed by poorer performance. Although participants changed roles based on their evolving preferences for how work should be divided, these changes did not necessarily lead to better task outcomes.

One possible explanation is that users may not always know which division of work will maximize performance, particularly when their perceptions of task difficulty or AI capability are imperfect~\cite{gor2026, fugener2019}. 

These results indicate that flexible role selection alone may be insufficient for effective human--AI collaboration in sequential decision-making tasks, as humans are poor at allocating subtasks correctly and the allocation process can impose additional cognitive overhead. Participants described this meta-decision as ``double thinking'' under time pressure, since they had to reason both about the task and about which collaborator should perform each subtask. Mixed-initiative systems could reduce this burden through lightweight suggestions or defaults, information about the relative strengths of the human and AI, or cues about uncertainty, while leaving the final allocation decision with the user.
\subsection{Behavioral signals can support role adaptation} 

Our analyses indicate that behavioral and task-specific signals contain enough information to predict whether users are likely to switch roles. In particular, gaze-related features contributed strongly to predictive performance, suggesting that implicit behavioral traces can help identify moments when users are reconsidering the current division of work. This extends prior work on adaptive human--AI systems, which often relies on system-side signals such as confidence, workload, or task state~\cite{fiore2016, salikutluk_evaluation_2024, hauptman2024}, by showing that user's own behavior can provide an additional source of information about changing role preferences.

Anticipating likely role changes creates an opportunity for timely, context-sensitive support. Our predictive model results suggests that behavioral and task-specific signals can identify moments when a user may be approaching a role change. Rather than automatically changing roles, behavioral predictions could be used to surface alternative allocations or assistance precisely when users appear to be reconsidering how work should be divided. Prior work has shown that system-initiated changes in task allocation, particularly when poorly time or misaligned with user's preferences, can reduce their sense of agency or trust in AI~\cite{luo2026, hauptman_adapt_2023}.

\section{Limitations and Future Work}

While our study offers insights into dynamic role switching in human–AI collaboration during a sequential decision-making task, several limitations should be noted. The small sample size and limited number of switching events per participant limit the generalizability and predictive strength of our model. Although we incorporated behavioral and contextual features, the model does not account for factors like evolving user strategy or individual trust calibration, which likely influenced switching decisions. Our study captured role switches within a single game session. However, user preferences may shift with extended use as participants adapt to different roles, refine their strategies, or recalibrate expectations based on growing familiarity with the AI's performance. 
Future work should therefore examine role-preference dynamics across multiple sessions and larger samples. Longitudinal studies could reveal how role-selection strategies, trust, and perceptions of AI alignment shift as users develop sustained experience with specific AI partners. Beyond longer-term studies, future work could extend our model to larger samples and other high-stakes sequential decision-making domains, such as clinical triage or trading. Building on our predictive results, future systems could investigate user-centered interventions such as context-sensitive role suggestions, capability or uncertainty cues, and intent explanations to determine whether anticipating role changes can improve collaboration without overriding user choice. Finally, temporally aware models that capture how behavioral and task signals evolve during deliberation may provide richer representations of the processes preceding role changes.

\section{Conclusion}

In this work we investigated how behavioral and task-specific signals relate to and can predict user-initiated role switching during sequential human-AI collaboration. Using a modified hand-and-brain chess paradigm, we found that role switches are shaped by perceived cognitive effort, decision risk, and changing assessments of the AI partner's competence, while gaze patterns provided a measurable signal prior to the observed switches. Our predictive model showed that behavioral and task-specific features can anticipate role changes at the subtask level. By identifying behavioral markers of role switching and demonstrating the feasibility of real-time prediction, this work provides a foundation for designing mixed-initiative systems that support timely, user-centered role adaptation while preserving user's agency over subtask allocation.


\section{Acknowledgments}
\paragraph{Chess.com Collaboration Disclaimer}: anonymized

\section{GenAI Usage Disclosure}
LLMs were used to explore potential data analysis approaches and to suggest recommendations for improving text phrasing and clarity. Any suggestions provided by LLMs served only as input to inform our own decision-making.

\bibliographystyle{ACM-Reference-Format}
\bibliography{chess}
\end{document}